\documentclass[superscriptaddress,reprint,twocolumn,aps,floatfix]{revtex4-2}
\usepackage[utf8]{inputenc}
\usepackage[T1]{fontenc}
\usepackage[english]{babel}
\usepackage{graphicx}
\usepackage{epsfig}
\usepackage{amsmath,amsfonts,amssymb}
\usepackage{color}
\usepackage{array}
\usepackage[loose]{units}
\usepackage{braket}
\usepackage{xr-hyper}

\usepackage[caption=false]{subfig}

\usepackage{physics}
\usepackage{soul}
\usepackage{upgreek}
\usepackage[colorlinks=true,citecolor=blue,linkcolor=magenta]{hyperref}
\usepackage{cleveref}

\graphicspath{{./Figures/}}

\newcommand{\MPINAT}{Max Planck Institute for Multidisciplinary Sciences, D-37077 G\"{o}ttingen, Germany}
\newcommand{\GOE}{University of G\"{o}ttingen, 4th Physical Institute, D-37077 G\"{o}ttingen, Germany}

\begin{document}


\title{Quantum eraser experiments for the demonstration of entanglement between swift electrons and light}

\author{Jan-Wilke Henke}
\affiliation{\MPINAT}
\affiliation{\GOE}

\author{Hao Jeng}
\affiliation{\MPINAT}
\affiliation{\GOE}

\author{Claus Ropers}
\email[]{claus.ropers@mpinat.mpg.de}
\affiliation{\MPINAT}
\affiliation{\GOE}

\date{\today}


\begin{abstract}
    We propose a tangible experimental scheme for demonstrating quantum entanglement between swift electrons and light, relying on coherent cathodoluminescence for photon generation in a transmission electron microscope, and a quantum eraser setup for formation and verification of entanglement.
    The entanglement of free electrons with light is key to developing free-electron quantum optics and its potential applications such as quantum sensing, novel photonic and electron state generation, and entanglement between free electrons.
\end{abstract}

\maketitle


\section{Introduction}

Entanglement between different subsystems or degrees of freedom is a defining hallmark of quantum science, and underpins unique applications in emerging quantum technologies such as quantum computation \cite{nielsen_quantum_2010, ladd}, communication \cite{gisin_quantum_2002,azuma,wehner} and sensing \cite{giovannetti_advances_2011, degen_quantum_2017}.
While entanglement can occur naturally by simply letting two quantum systems interact, it is also notoriously fragile and difficult to observe because the quantum correlations are easily overwhelmed by decoherence.

The quantum eraser provides a particularly striking and conceptually instructive demonstration of quantum entanglement \cite{ma_delayed-choice_2016}, in which eliminating the which-path information causes the recovery of multipath interference otherwise lost by creating an entangled marker particle along the path.
With the addition of suitable inseparability criteria \cite{horodecki, guehne, clauser_proposed_1969, aspect_experimental_1982, storz_loophole-free_2023}, one could verify if the two parties of the system exist in a state of entanglement.
Initially proposed as a Gedanken experiment, the quantum eraser has been demonstrated with photons \cite{rarity_experimental_1990,herzog_complementarity_1995,walborn_double-slit_2002, kim_delayed_2000, kaiser_entanglement-enabled_2012}, atoms \cite{durr_origin_1998}, electrical circuits \cite{weisz_electronic_2014}, and phonons \cite{bienfait_quantum_2020}.

Perhaps surprisingly, the possibility of entangling fast electrons, used in electron microscopy for research on nanoscale structures and dynamics \cite{yip_atomic-resolution_2020, baum_4d_2007, danz_ultrafast_2021}, has only recently begun to attract attention \cite{schattschneider_entanglement_2018, kfir_entanglements_2019, zhao_quantum_2021, konecna_entangling_2022}, despite their exceptional controllability and favorable coherence properties.
Spontaneous inelastic scattering of electrons is routinely employed in the study of optical excitations \cite{garcia_de_abajo_optical_2010}, and quantum optics has entered this field in the form of photon correlation spectroscopy \cite{meuret_photon_2015, sola-garciaPhotonStatisticsIncoherent2021, scheucherDiscriminationCoherentIncoherent2022, fiedler_sub--super-poissonian_2023}.
The stimulated inelastic interaction with optical near-fields is quantum coherent \cite{barwick_photon-induced_2009, feist_quantum_2015}, and has enabled optical field characterization \cite{piazza_simultaneous_2015, nabbenAttosecondElectronMicroscopy2023, yangFreeelectronInteractionNonlinear2024}, reconstruction of the free electron quantum state \cite{priebeAttosecondElectronPulse2017a} as well as free-electron homodyne detection \cite{gaida_Attosecond_2024}.
Numerous applications harnessing the quantum nature of this inelastic electron-light scattering and the resulting correlations have been suggested, including probing of quantum optical excitations \cite{di_giulio_probing_2019, dahan_imprinting_2021}, correlation-enhanced imaging \cite{feist_cavity-mediated_2022, varkentina_cathodoluminescence_2022}, improved measurement sensitivity in interaction-free measurements  \cite{kruit_Designs_2016, turner_interaction-free_2021, koppell_Transmission_2022, rotunno_one-dimensional_2023}, and the generation of novel quantum states of light \cite{bendana_single-photon_2011, ben_hayun_shaping_2021, dahan_creation_2023}.
But while the underlying interactions are expected to induce electron-photon entanglement \cite{feist_cavity-mediated_2022, konecna_entangling_2022}, facilitate electron-electron entanglement \cite{kfir_entanglements_2019} or even mediate photon-photon entanglement \cite{baranes_free_2022}, studies thus far have fallen short of direct proof.

The objective of this paper is to describe experimental scenarios for demonstrating the entanglement of free electrons and light.
For illustrative purposes, we first consider the suppression and subsequent recovery of single-electron interference resulting from quantum correlations in a quantum eraser scenario.
Specifically, we introduce a double-slit geometry producing entanglement between photonic degrees of freedom and the electron position (Sec. \ref{sec:quantum_eraser})
Addressing experimental implementations, in Section \ref{sec:exp_consideration}, we propose dual-point probes as used in STEM holography and coincidence measurements to generate optical excitations at designed photonic structures to form an entangled bipartite state and perform characterising measurements.
Finally, we relate the measurements in this quantum eraser scenario to entanglement tests such as quantum state tomography of the electron-photon system (Sec. \ref{sec:tomography}), and discuss a transfer also to free electron-electron entanglement.


\section{Main}

\subsection{Concept of quantum erasure}
\label{sec:quantum_eraser}

\begin{figure*}[ht]
    \centering
    \includegraphics[width=0.95\textwidth]{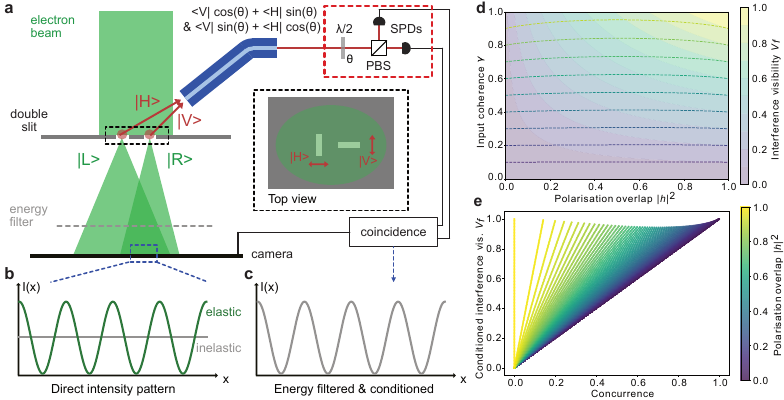}
    \caption{\textbf{Quantum eraser experiment with free electrons.}
            \textbf{a} An electron beam illuminates a double-slit structure. Electrons (green) passing through the left (quantum state $\ket{L}$) and right ($\ket{R}$) slit generate identical photons (red) of orthogonal polarisations $\ket{V}$ and $\ket{H}$ as highlighted in the inset. These marker photons are collected with a fiber (blue), passed through a half-wave plate ($\lambda/2$) of variable orientation $\theta$ and a polarising beam splitter (PBS) before detection on single-photon detectors (SPDs), thus, realising projective measurements on superposition of $\ket{H}$ and $\ket{V}$ (dashed red box). The electrons are energy-filtered for one-photon loss (dashed grey line) and detected on a camera, enabling coincidence detection of electrons and photons.
            \textbf{b} Direct intensity distribution behind the energy filter exhibiting interference fringes when no marker photon is generated (green) and no interference (grey) when orthogonal marker photons are generated.
            \textbf{c} Interference pattern recovered from coincidences of energy-filtered electrons and photons after local operations with the wave plate.
            \textbf{d} Visibility $V_f$ of the interference fringes depending on the input coherence $\gamma$ and the overlap of the marker photons $|h|^2$ in the direct (colour map \& solid lines) and conditioned intensity distribution (dashed lines).
            \textbf{e} Interference fringe visibility $V_f$ in the recovered distribution versus the concurrence $C$ of the electron-photon state for different polarsation overlap $|h|^2$ (colour coded) and input coherences.
            }
    \label{fig:1}
\end{figure*}

The basic idea of a quantum eraser relies on single-particle interference observable behind a double-slit structure.
When introducing a marker, entangled with the interfering particle and providing which-path information, the inferference disappears but can be recovered using a basis change on the marker and coincidence detection \cite{ma_delayed-choice_2016}. 
These concepts can be applied to experiments with free electrons, as illustrated in Figure \ref{fig:1}(a). 

When a fully coherent electron beam homogeneously illuminates a double-slit structure, interference causes an oscillation in the intensity distribution in the far field \cite{jonsson_elektroneninterferenzen_1961, bach_controlled_2013, tavabi_young-feynman_2019}.
The intensity pattern, given by $I(x) \approx I_0(x) (1 + \cos(\phi(x)))$ with the diffraction pattern of a single slit $I_0(x)$ and the phase difference between the propagation pathways $\phi(x)$ (see section \ref{sec:appendix} for details), can be detected using a camera as shown in Figure \ref{fig:1}(b).

Suppose that electrons passing through the slits generate distinguishable photons, e.g. in different spatial modes or of orthogonal polarisation.
For simplicity, assume that transmission through the left (right) slit gives horizontally (vertically) polarized photons, denoted $\ket{H}$ and $\ket{V}$ respectively.
This results in an entangled electron-photon state
\begin{align}
    \ket{\psi} = \frac{1}{\sqrt{2}} (\ket{L, H} + \ket{R, V})
\label{eq:max_entangled}
\end{align}
behind the slit with the generated photons carrying which-path information about the electron, eliminating the electron interference pattern (grey line in Fig. \ref{fig:1}(b).
An electron energy filter (dashed gry line in Fig. \ref{fig:1}(a)) selects the fraction of electrons that produced a marker photon and lost the corresponding energy.

The interference pattern can, however, be restored with a photon state basis change and a projective measurement.
To this end, the marker photons are collected via an optical single-mode fiber and passed through a half-wave plate as well as a polarising beam splitter (red box in Fig. \ref{fig:1}(a)).
The wave plate effectively erases the which-path information by mixing the polarisation states rendering the single-photon detectors (SPDs) placed behind a polarising beam splitter (PBS) unable to distinguish the electron paths.
This results in a recovered interference pattern from coincident photons and energy-filtered electrons as shown in figure \ref{fig:1}(c).

\subsection{Experimental considerations}
\label{sec:exp_consideration}

Under realistic experimental conditions imperfections in the marker photon generation process, the electron beam preparation or the optical setup will hamper the elimination and coincidence-based recovery of the interference pattern.
A deviation from perfectly orthogonal marker photon states, for example the case of electrons passing through the right slit generating a photon in a superposition of polarisations $h \ket{H} + v \ket{V}$, results in the unconditioned $(I_u)$ and recovered, conditioned intensity patterns ($I_c$): 
\begin{align*}
    I_u(x) &= I_0(x) + \text{Re}(h) I_0(x) \cos(\phi(x))~,
\end{align*}
and
\begin{align*}
    I_c(x) &= (1 + \text{Re}(h^* v)) I_0(x) + \text{Re}(h + v) I_0(x) \cos(\phi(x))~.
\end{align*}
Notably, the unconditioned intensity distribution exhibits oscillations depending on the overlap $|h|^2$ of the marker photon states.
A larger overlap reduces the which-path information about the electron and is directly linked to a reduced degree of electron-photon entanglement, which can be quantified via the concurrence $C$ \cite{plenio_introduction_2007, guehne}.
For our case of a pure bipartite state, it can be expressed as $C = |h|$, taking on values between $0 \leq C \leq 1$ depending on the polarisation overlap.

Any principal distinguishability of the marker photons in other degrees of freedom, such as the wavelength, will reduce the visibility of the interference fringes $V_f = (I_{max} - I_{min})/(I_{max} + I_{min})$ due to the reduced overlap in the partial trace. 
Careful design of both the sample as well as the optical setup is, therefore, paramount to avoid this in the generation and propagation of the photons.

Similar care needs to be taken in the preparation of the electron beam, as limited spatial coherence $\gamma < 1$ will suppress the off-diagonal elements of the bipartite state's density matrix and, thus, impose an upper bound on the concurrence.
This is accompanied by a reduction of visibility in both the unconditioned and recovered interference patterns.

These findings are summarised in figure \ref{fig:1}(d)\&(e), where panel d shows the dependence of the fringe visibility $V_f$ in the direct pattern (contours and solid) and reconstructed, conditioned pattern (dashed lines) on the polarisation state overlap $|h|^2$ and input beam coherence $\gamma$.
The former increases with the overlap $|h|^2$ and is bounded by the degree of coherence, while the latter exhibits only small variations.
When comparing the recovered interference fringe visibility $V_f$ with the concurrence $C$ of the electron-photon state within our simple model (see Fig. \ref{fig:1}(e)), we find an almost perfect linear relation for small polarisation overlap $|h|^2 < 0.1$.
In these cases, quantum erasure, signalled by the recovery of interference fringe visibility after conditioning, can be considered an indicator of electron-photon entanglement.


An experimental realisation of free electron-photon quantum erasure, accordingly, requires:
First, a highly coherent and controllable electron beam to illuminate a double slit-type structure.
Second, efficient generation of distinct marker photons carrying the electron which-path information in a single degree of freedom.
Third, the capability to collect marker photons, manipulate their quantum state manipulation and perform projective measurements in coincidence with energy-filtered electron detection in the diffraction pattern.

\begin{figure*}[ht]
    \centering
    \includegraphics[width=0.95\textwidth]{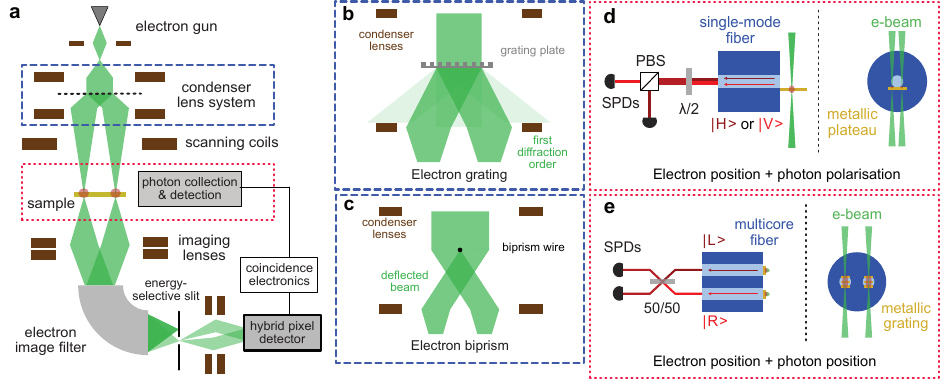}
    \caption{\textbf{Possible implementations of a quantum eraser experiment.}
            \textbf{a} Key elements of a free-electron quantum eraser are electron beam splitting, efficient marker photon generation and collection, and coincidence detection.
            Splitting the electron beam in the condenser system (blue box, details in b\&c) two focused electron probes are formed at the sample.
            There, distinguishable marker photons are generated and collected for further processing (red box, details in c-f).
            Below the sample, an imaging filter selects electrons that generated a marker photon before recombining the beams to give an interference pattern on a hybrid-pixel detector that enables coincidence detection of electrons and photons.
            \textbf{b} The splitting of the electron beam can be achieved, for example, by a grating plate causing diffraction of the electron beam.
            \textbf{c} Alternatively, an electron biprism consisting of a thin, biased wire deflecting parts of the input beams in different directions can be used.
            \textbf{d} The generation of polarised photons ($\ket{H}$ or $\ket{V}$), e.g. in transition radiation, at a specifically designed structure placed in front of an optical fiber allows for marker-photon generation and collection. 
            Manipulation of the which-path information is then achieved by a waveplate ($\lambda / 2$) and a polarising beam splitter (PBS) in front of the single-photon detectors (SPDs).
            \textbf{e} Alternatively, marker photons can be generated via Smith-Purcell radiation coupled to different fiber cores, yielding photons in the states $\ket{L}$ or $\ket{R}$. An erasure of the which-path information can then be implemented using a balanced beam splitter ($50/50$) before the SPDs.
            }
    \label{fig:2}
\end{figure*}

Transmission electron microscopes (TEMs), particularly those with field-emission electron sources, provide a well-controlled and coherent electron beam, enabling, e.g., electron holography \cite{tonomura_applications_1987, midgley_electron_2009}. 
Splitting the electron beam, as illustrated in figure \ref{fig:2}(a), rather than using a transmissive double slit structure slightly relaxes the coherence requirements and significantly increases the effective electron current.
The splitting can, for example, be achieved by deflection using an electrostatic biprism \cite{tonomura_applications_1987, tavabi_young-feynman_2019} or diffraction from a holographic phase or amplitude plate in the TEM's condenser system \cite{r_harvey_efficient_2014, yasin_path-separated_2018}.
Amplitude plates, schematically depicted in figure \ref{fig:2}(b), require careful design and fabrication to ensure low losses and a deflection of electrons mainly into the first diffraction order, whereas a biprism (c.f. Fig. \ref{fig:2}(c)) allows tunable separation of beams but demands higher beam coherence as different sections of the beam need to interfere. 
In both cases, the beams must be overlapped below the sample to observe interference (cf. Fig. \ref{fig:2}(a)), either through the imaging lenses or additional beam splitting elements in a Mach-Zehnder interferometer scheme \cite{johnson_scanning_2021}.

Distinguishable marker photons for the two electron pathways can be generated via different coherent parametric processes \cite{garcia_de_abajo_optical_2010,christopher_electron-driven_2020}, including inelastic electron-light scattering, Smith-Purcell radiation or transition radiation.
Two possible geometries allowing for the entanglement of the electron position with different photonic degrees of freedom are presented in figure \ref{fig:2}(d)\&(e).

Our first suggestion (Fig. \ref{fig:2}(d)) involves polarised photon generation at a specifically shaped metallic structure placed in front of a single optical fiber.
Fast electrons impinging on a metallic surface generate transition radiation due to the annihilation of the image charge in the material, leading to the emission of polarized electromagnetic radiation \cite{garcia_de_abajo_optical_2010}.
Tailoring the boundary properties, electrons hitting the metallic plateau at different positions relative to the fiber core will generate different photon polarizations.
At typical TEM energies, the photon generation efficiencies are on the order of $10^{-3}$, and the emission is broadband \cite{garcia_de_abajo_optical_2010}, rendering the non-orthogonality of marker photons the main concern for the recovered electron fringe visibility.
The required manipulation of the photon state and erasure of which-path information can be implemented using wave plates, a PBS and SPDs as indicated in figures \ref{fig:1}(a) and \ref{fig:2}(d).

Alternatively, one could achieve entanglement of the electron path and the generated photon position at the endface of a multicore fiber (c.f. Fig. \ref{fig:2}(e)) or the photon propagation direction at an optical waveguide.
In the former case, grating structures imprinted on metal-coated fiber endfaces facilitate photon generation in the different fiber cores $\ket{L}$ or $\ket{R}$ via the Smith-Purcell effect \cite{so_fiber_2014, so_amplification_2015}.
The electrons' evanescent field is diffracted into the fiber with the emission wavelength and direction determined by the grating parameters and the electron velocity \cite{garcia_de_abajo_optical_2010, christopher_electron-driven_2020}.
The latter scenario uses inelastic scattering of free electrons at the vacuum field of a well-defined resonator mode \cite{kfir_entanglements_2019, muller_broadband_2021}, with phase-matched interactions shown to enable the generation of correlated electron-cavity photon pairs in integrated photonic circuits \cite{feist_cavity-mediated_2022}.
In both cases, erasure of the electron which-path information can be achieved by a balanced beam splitter mixing the two photon pathways.
Despite higher potential photon generation efficiencies, these scenarios require larger electron beam separations and precise sample design to ensure required marker photon properties, thereby posing additional experimental challenges.
Other experimental geometries, e.g. based on parabolic mirrors for free-space photon collection \cite{meuret_photon_2015, varkentina_cathodoluminescence_2022}, and harnessing different coherent cathodoluminescence processes are also possible.

An energy-filtering electron imaging spectrometer, illustrated at the bottom of figure \ref{fig:2}(a), can select electrons associated with a marker photon by the corresponding energy-loss \cite{pomarico_mev_2018, henke_integrated_2021}.
Combined with a hybrid-pixel electron detector, this allows for a time-, energy- and position-resolved detection of the electrons with high signal-to-noise ratio, curcial for coincidence-based measurements \cite{feist_cavity-mediated_2022, varkentina_cathodoluminescence_2022}.
Coincidence detection of loss-electrons and photons, using an event-based detector and SPDs performing the projective measurement, then enables the recovery of the electron interference pattern.

\subsection{Tests for inseparability}
\label{sec:tomography}

Measurements performed with the quantum eraser setup can also more generally be used to determine whether the electron-light system is entangled.
Let us denote the Pauli matrices by $\sigma_x$, $\sigma_y$, $\sigma_z$ as usual, and associate the photon and electron states \{$\ket{H}$, $\ket{V}$\} and \{$\ket{L}$, $\ket{R}$\} with the eigenvectors of $\sigma_z$.
Then, the fidelity of any given state with respect to the Bell state in equation \eqref{eq:max_entangled} can be expressed as $F = (1 + \expval{\sigma_x \otimes \sigma_x} - \expval{\sigma_y \otimes \sigma_y} + \expval{\sigma_z \otimes \sigma_z})/4$, and a value greater than one-half provides a sufficient condition for entanglement \cite{guehne}.
In our quantum eraser setup, these correlation functions can be obtained by first measuring the photon in the respective basis (i.e. $\sigma_x$, $\sigma_y$, or $\sigma_z$), then summing up the visibilities of the electron interference fringes observed when conditioning on each of the photon states in that basis.
Not all situations where interference fringes are recovered correspond to entanglement, however, just those where the visibilities are large enough that the inequality above is satisfied.

Interestingly, tomographic reconstruction of the two-qubit electron-photon density matrix is also possible, since all independent combinations of Pauli measurements can be implemented.
Consequently, it is straightforward to determine whether the state is entangled by using entanglement measures \cite{horodecki}.
On the other hand, very little information is needed to settle the question of separability, and complete quantum tomography is certainly not necessary.
For instance, the criterion $F>1/2$ can be relaxed to include only measurements of $\sigma_x$ and $\sigma_z$, becoming $|\expval{\sigma_x \otimes \sigma_x} + \expval{\sigma_z \otimes \sigma_z}| > 1$ at the expense of requiring slightly higher correlations; this simplified measurement scheme corresponds to the most common description of the quantum eraser experiment.


\section{Conclusion and Outlook}

We have described conceptually simple quantum eraser experiments based on free electrons and light that appear within reach of current technology, to generate electron-light entanglement and verify that they are indeed entangled.
An especially interesting application of this scheme concerns the generalisation to \textit{electron-electron} entanglement, by creating two sets of such electron-photon entangled states and heralding on the detection of two photons with orthogonal polarizations $\ket{H}$ and $\ket{V}$.
The measurement erases all information possessed by the photons about which electron went through which slit, and projects the two electrons into an entangled state $(\ket{L}\ket{R} + \ket{R}\ket{L})/\sqrt{2}$.
If for any reason the two photons originate from different locations, the same effect can be achieved by performing a Bell-state measurement on the photons, thereby swapping the entanglement from the photon of one electron-photon pair to the electron of the other pair.
The principles of quantum metrology strongly suggest that entanglement will be useful for improving the sensitivity of electron microscopes.
The large probability of damage caused by irradiation with swift electrons in electron microscopy makes the creation of entanglement involving free electrons all the more important.

\section*{Acknowledgements}
The authors would like to thank Murat Sivis, Ofer Kfir, Armin Feist, Germaine Arend, Philipp Haslinger, Valerio Di Giulio and Javier Garcia de Abajo for useful discussions.
This work was funded by the Deutsche Forschungsgemeinschaft (DFG, German Research Foundation) through grant number 432680300/SFB 1456 (project C01) and the Gottfried Wilhelm Leibniz programme, and the European Union’s Horizon 2020 research and innovation programme under grant agreement number 101017720 (FET-Proactive EBEAM).


\section{Appendix}
\label{sec:appendix}

\subsection{Quantum eraser}

This sections aims to provide a more detailed derivation of the electron intensity distributions described in the main text sections \ref{sec:quantum_eraser} \& \ref{sec:exp_consideration}.

The intensity distribution for electrons on the screen can be calculated via: 
\begin{align*}
    I(x) &= \rho(x,x) = |a|^2 |\bra{x}U\ket{L}|^2 + |b|^2 |\bra{x}U\ket{R}|^2 \\
    &+ a b^* \bra{x}U\ket{L}\bra{R}U^{\dagger}\ket{x} + a^* b \bra{x}U\ket{R}\bra{L}U^{\dagger}\ket{x}~,
\end{align*}
where $\rho(x,x) = \bra{x} \rho \ket{x}$ denotes the diagonal terms of the density matrix in position representation and $U$ is the propagator for the electrons from the slit plane with the left and right slit (denoted $\ket{L}$ or $\ket{R}$, respectively) to the point $x$ in the detector plane.
At short distances $U$ can be approximated by the Fresnel propagator \cite{lubk_chapter_2018}, while in the far-field this reduces to Fourier transforms $\mathcal{F}$ of the wavefunction in the slit plane with coordinate $x'$:
\begin{align*}
    I(x) &= |a|^2 |\mathcal{F}(\braket{x'}{L})|^2 + |b|^2 |\mathcal{F}(\braket{x'}{L})|^2 \\
    &+ a b^* e^{- i \phi(x)} \mathcal{F}(\braket{x'}{L}) \mathcal{F}^*(\braket{x'}{R})\\
    &+ a^* b e^{i \phi(x)} \mathcal{F}(\braket{x'}{R}) \mathcal{F}^*(\braket{x'}{L}) \\
    &= I_0(x) + I_0(x) \cos(\phi(x))~,
\end{align*}
where $\phi(x)$ is the phase difference accumulated between electrons from the two slits to a point $x$ on the detector and $I_0(x)$ describes the single slit diffraction pattern, assuming evenly illuminated ($a = b$), identical slits.

Now consider the case of the electron generating distinguishable marker photons.
With the electron passing through the left (right) slit generating a photon of polarisation $\ket{H}$ ($\ket{V}$), the electron-photon density matrix behind the slit may be written as:
\begin{align*}
    \rho &= |a|^2 \ket{L, H}\bra{L, H} + |b|^2 \ket{R, V}\bra{R, V} \\
    &+ a b^* \ket{L, H}\bra{R, V} + a^* b \ket{R, V}\bra{L, H}~,
\end{align*}
resulting in an electron intensity distribution on the camera given by:
\begin{align*}
    I_m(x) &= \bra{x} \Tr_{ph} (U \rho U^{\dagger}) \ket{x} \\
    &= |a|^2 |\mathcal{F}(\braket{x'}{L})|^2 + |b|^2 |\mathcal{F}(\braket{x'}{L})|^2\\
    &= I_0(x)~.
\end{align*}
with $\Tr_{ph} (\rho)$ being the partial trace over the photon degrees of freedom and the last line again assuming even illumination of the double slit.
Clearly, the entanglement induced by the generation of the marker photons removes the interference pattern as the photons carry which path information about the individual electrons.

Collecting the generated marker photons and passing them through a half-wave plate (c.f. Fig. \ref{fig:1}(a)) transforms the polarization states in the following manner: $\ket{H} \longrightarrow (\ket{H} + \ket{V}) / \sqrt{2}$ and $\ket{V} \longrightarrow (\ket{H} - \ket{V})/\sqrt{2}$.
Placing a polarising beam splitter in the beam path that separates the orthogonal polarizations $\ket{H}$ and $\ket{V}$ and conditioning on the detection of a photon with polarization $\ket{H}$ (with identity operator $\text{id}$ on the electron side), one finds the modified state:
\begin{align*}
    \Tilde{\rho} &= \frac{(\text{id} \otimes \ket{H}\bra{H}) \rho (\text{id} \otimes \ket{H}\bra{H})}{\Tr{\rho (\text{id} \otimes \ket{H}\bra{H})}} \\
    &= |a|^2 U \ket{L, H}\bra{L, H} U^{\dagger} + |b|^2 U \ket{R, H}\bra{R, H} U^{\dagger} \\
    &+ a b^* U \ket{L, H}\bra{R, H} U^{\dagger} + a^* b U \ket{R, H}\bra{L, H} U^{\dagger}~,
\end{align*}
resulting in the intensity distribution:
\begin{align*}
    I_e(x) 
    &= |a|^2 |\mathcal{F}(\braket{x'}{L})|^2 + |b|^2 |\mathcal{F}(\braket{x'}{L})|^2 \\
    &+ a b^* e^{-i \phi(x)} \mathcal{F}(\braket{x'}{L}) \mathcal{F}^*(\braket{x'}{R})\\
    &+ a^* b e^{i \phi(x)} \mathcal{F}(\braket{x'}{R}) \mathcal{F}^*(\braket{x'}{L})\\
    &= I_0(x) + I_0(x) \cos(\phi(x))~,
\end{align*}
where the interference fringes are recovered.
Analogously, post-selecting photons of polarization $\ket{V}$ behind the half-wave plate results in an interference pattern with inverted minima and maxima due to a sign change in the last two terms.

More generally, if the marker photons are not perfectly orthogonal, i.e. the post interaction state reads $\ket{\psi}_D = a \ket{L, H} + b \ket{R} \otimes (h \ket{H} + v \ket{V})$, both the unconditioned and the path-information erased, conditioned intensity distributions ($I_u$ and $I_c$, respectively) change:
\begin{align*}
    I_u(x) &= |a|^2 |\mathcal{F}(\braket{x'}{L})|^2 + |b|^2 |\mathcal{F}(\braket{x'}{L})|^2  \notag \\ 
    &+ a b^* h^* e^{-i \phi(x)} \mathcal{F}(\braket{x'}{L}) \mathcal{F}^*(\braket{x'}{R}) \notag \\
    &+ a^* b h e^{i \phi(x)} \mathcal{F}(\braket{x'}{R}) \mathcal{F}^*(\braket{x'}{L}) \\
    &= I_0(x) + \text{Re}(h) I_0(x) \cos(\phi(x))~,
\end{align*}
and
\begin{align*}
    I_c(x) &= |a|^2 |\mathcal{F}(\braket{x'}{L})|^2 + |b|^2 (1 + h^* v + h v^*) |\mathcal{F}(\braket{x'}{L})|^2 \notag \\ 
    &+ a b^* (h^* + v^*) e^{-i \phi(x)} \mathcal{F}(\braket{x'}{L}) \mathcal{F}^*(\braket{x'}{R}) \notag \\
    &+ a^* b (h + v) e^{i \phi(x)} \mathcal{F}(\braket{x'}{R}) \mathcal{F}^*(\braket{x'}{L}) \\
    &= (1 + \text{Re}(h^* v)) I_0(x) + \text{Re}(h + v) I_0(x) \cos(\phi(x))~. 
\end{align*}
The unconditioned intensity distribution $I_u(x)$ exhibits oscillations for $h \neq 0$ as the marker photons are no longer perfectly distinguishable.
At the same time, the impact on the conditioned intensity distribution $I_c(x)$ reconstructed from the electron-photon coincidences after manipulation is less pronounced.
Both the unconditioned and the recovered, conditioned intensity patterns can be characterised by a corresponding visibility $V_f$ of the interference fringes, shown in figure \ref{fig:1}(d).

The description of the quantum eraser process can be modified to account for limited spatial coherence by replacing the electron state below the slit with the mixed state:
\begin{align*}
    \rho &= \gamma \ket{\psi}\bra{\psi} + (1 - \gamma) (|a|^2 \ket{L, H}\bra{L, H} \\
    &+ |b|^2 \ket{R}\otimes(h \ket{H} + v \ket{V}) (h^* \bra{H} + v^* \bra{V}) \otimes \bra{R})~,
\end{align*}
where $\gamma$ is linked to the degree of transverse coherence.
In this mixed state case the the concurrence $C$ is defined as $C(\rho) = \max(0, \lambda_1 - \lambda_2 - \lambda_3 - \lambda_4)$ where the $\lambda_i$ denote the eigenvalues, in decreasing order, of $R = \sqrt{\sqrt{\rho} \tilde{\rho} \sqrt{\rho}}$ with $\tilde{\rho} = (\sigma_2 \otimes \sigma_2) \rho^* (\sigma_2 \otimes \sigma_2)$ and the Pauli matrix $\sigma_2$.
Figure \ref{fig:1}(e) presents the relation between the concurrence and the visibility of the recovered, conditioned intensity pattern for this simple model.

\subsection{Quantum state tomography}
This section aims to provide background information to section \ref{sec:tomography}.
The electron-photon quantum state in the proposed experimental scenario is equivalent to a two-qubit state and can be represented in the form:
\begin{widetext}
\[
    \rho = a \text{id} \otimes \text{id} + \sum_{i = x,y,z} (b_i \sigma^e_i \otimes \text{id} + c_i \text{id} \otimes \sigma^{ph}_i) + \sum_{i,j = x,y,z} d_{ij} \sigma^e_i \otimes \sigma^{ph}_j~,
\]
\end{widetext}
with $\sigma_k$ for $k=x,y,z$ as the three Pauli matrices as usual, and superscripts $e$ and $ph$ to emphasize whether we are referring to the electron or photon. We take as eigenstates of $\sigma_z$ the basis $\{\ket{L}, \ket{R}\}$ for the electron and $\{\ket{H}, \ket{V}\}$ for the photon.
The coefficients are determined by local measurements of the corresponding observables: 
\begin{align*}
    b_i &= \Tr((\sigma_i \otimes \text{id}) \rho),\\
    c_i &= \Tr((\text{id} \otimes \sigma_i) \rho),\\
    d_{ij} &= \Tr((\sigma_i \otimes \sigma_j) \rho),
\end{align*}
and with $a$ fixed by normalization. The density matrix of the two-qubit system could be completely determined as long as one could implement measurements in all three bases independently for the electron and for the photon.

On the photon side, measurements of the three Pauli observables correspond to projections onto the different polarization states: horizontal or vertical ($\sigma_z$), diagonal ($\sigma_x$), and left and right-handed circular polarisations ($\sigma_y$), and these can be implemented in the usual way by using half-waveplates, quarter-waveplates, polarising beam-splitters, and avalanche photon detection.
On the electron side, the situation is more complicated but, nevertheless, measurements in all three bases can be implemented. A real-space image of the two slits obtained in the conventional imaging mode of the microscope tells us which slit the electron goes through, thus realising a measurement of $\sigma_z$.
Diffractograms obtained in the diffraction mode correspond to interference between the two beams with some relative phase shift $\phi(x)$, with $x$ denoting the position on the screen, so that the detection of an electron at a position $x$ on the screen corresponds to the projection operator $\bra{L} + e^{i\phi(x)}\bra{R}$.
Therefore, measurements in the diffraction plane allow us to measure $\sigma_x$ and $\sigma_y$. Complete knowledge of the density matrix can then be used to evaluate entanglement measures such as the concurrence $C$ or the negativity \cite{horodecki}.

Alternatively, the separability of the state could also be ascertained through fewer measurements by employing entanglement witnesses \cite{horodecki,guehne}.
The joint electron-photon quantum state is ideally represented by the maximally entangled state $\ket{\psi} = (\ket{L}\ket{H} + \ket{R}\ket{V})/\sqrt{2}$, thus we should consider the fidelity between this state and an experimentally generated state $\rho$:
\begin{align*}
    F &= \expval{\psi | \rho | \psi} \\
      &=  \frac{1}{4} \Tr(\rho (1 + \sigma^e_x \otimes \sigma^{ph}_x - \sigma^e_y \otimes \sigma^{ph}_y + \sigma^e_z \otimes \sigma^{ph}_z)),
\end{align*}
where the expression in the second line follows from the expansion of the state $\ket{\psi}\bra{\psi}$ in the Pauli operators.
Expectation values of outer products of Pauli operators can be measured as described above and a value of $F>1/2$ is sufficient for entanglement given the entanglement witness $\mathcal{W} = \text{id} - \ket{\psi}\bra{\psi}$ \cite{guehne}.
The expectation value of $\mathcal{W}$ can be expressed as $\langle \mathcal{W} \rangle = \frac{1}{2} - F(\rho, \ket{\psi}\bra{\psi})$, so that $\expval{W} < 0$ is equivalent to $F>1/2$.
By applying the triangle inequality, the inseparability criterion above could be further reduced to:
\begin{align*}
    |\expval{\sigma^e_x \otimes \sigma^{ph}_x} + \expval{\sigma^e_z \otimes \sigma^{ph}_z}| > 1 ~,
\end{align*}
which describes the standard quantum eraser scenario with measurements in two linearly polarised bases.


\bibliographystyle{apsrev4-2}
\bibliography{QE_arxiv}{}

\end{document}